\documentclass{elsart3p}

\usepackage{graphicx}
\usepackage{multicol}

\begin{document}

\begin{frontmatter}

\title{Unconventional pairing in bipolaronic theories}

\author[a]{J.P.Hague}
\author[b]{P.E.Kornilovitch}
\author[a]{A.S.Alexandrov}
\author[a]{J.H.Samson}

\address[a]{Department of Physics, Loughborough University, Loughborough, LE11 3TU, UK}
\address[b]{Hewlett-Packard Company, 1000 NE Circle Blvd, Corvallis, Oregon 97330, USA}

\begin{abstract}
Various mechanisms have been put forward for cuprate
superconductivity, which fit largely into two camps: spin-fluctuation
and electron-phonon (el-ph) mechanisms. However, in spite of a large effort,
electron-phonon interactions are not fully understood away from
clearly defined limits. To this end, we use a numerically exact
algorithm to simulate the binding of bipolarons. We present the
results of a continuous-time quantum Monte-Carlo (CTQMC) algorithm on a
tight-binding lattice, for bipolarons with arbitrary interaction range in the presence of strong coulomb repulsion. The algorithm is sufficiently efficient
that we can discuss properties of bipolarons with various pairing
symmetries. We investigate the effective mass and binding energies of
singlet and triplet real-space bipolarons for the first time, and discuss the extensions necessary to investigate $d$-symmetric pairs.
\end{abstract}

\begin{keyword}
% keywords here, in the form: keyword; keyword
Electron-phonon interactions\sep Strong correlations\sep Bipolaron superconductivity
\end{keyword}
\end{frontmatter}

\section{Introduction}

The discovery of superconductivity at anomalously high temperatures in cuprate materials caused a major revitalisation in the twin fields of correlated electrons and superconductivity. Such large transition temperatures could not be explained consistently within the BCS approach, which is only accurate in the extreme adiabatic, weak coupling limit (see e.g. refs \cite{hague2006a}). Several mechanisms have been put forward to understand these compounds, including very strong el-ph coupling, with bipolarons forming a Bose-Einstein condensate (BEC) in the finite density limit, leading to a bipolaronic superconductor (BPS) \cite{alexandrovmott}.

BPS theory and other approaches such as extended Eliashberg theory (EET)  \cite{hague2006a} are applicable in certain limits, either strong coupling (BPS) or weak couping (BCS and EET). However, the cuprates (and other strongly correlated electron systems) often have intermediate coupling between electrons. Thus, one cannot expand about a well defined solution, and it is necessary to employ advanced simulation techniques \cite{kornilovitch,hague,condmat}. Such techniques have the power to compute the numerically exact solution to a problem, and can provide definitive answers about the limits of validity of approximate schemes.

Here, we simulate a Coulomb-Fr\"{o}hlich (CF) Hamiltonian \cite{coulombfrohlich} with two electrons of opposite spin,
\begin{eqnarray}
H & = & - t \sum_{\langle \mathbf{nn'} \rangle\sigma}
c^{\dagger}_{\mathbf{n'}\sigma} c_{\mathbf{n}\sigma} + \sum_{
\mathbf{nn'}\sigma}V(\mathbf{n},\mathbf{n}') c^{\dagger}_{\mathbf{n}\sigma}
c_{\mathbf{n}\sigma}c^{\dagger}_{\mathbf{n'}\bar{\sigma}} c_{\mathbf{n'}\bar{\sigma}} \nonumber\\ &+& 
\sum_{\mathbf{m}} \frac{\hat{P}^{2}_\mathbf{m}}{2M} + 
\sum_{\mathbf{m}} \frac{\xi^{2}_{\mathbf{m}} M\omega^2}{2} -
\sum_{\mathbf{n}\mathbf{m}\sigma} f_{\mathbf{m}}(\mathbf{n})
c^{\dagger}_{\mathbf{n}\sigma} c_{\mathbf{n}\sigma} \xi_{\mathbf{m}}\nonumber
\: . \label{eq:four}
\end{eqnarray}
 Each ion has a displacement $\xi_\mathbf{m}$. Sites labels are $\mathbf{n}$ or $\mathbf{m}$
for electrons and ions respectively. $c$ annihilate
electrons. The phonons are Einstein oscillators with
frequency $\omega$ and mass $M$.
$\langle\mathbf{n}\mathbf{n}'\rangle$ denote pairs of nearest
neighbours, and $\hat{P}_{\mathbf{m}}=-i\hbar\partial/\partial\xi_{\mathbf{m}}$ ion momentum operators. The instantaneous interaction
$V(\mathbf{n},\mathbf{n}')$ has on-site repulsion $U$ and a n.n. attraction $V$ ($UV$ model). The Fr\"ohlich interaction force is long-range,
$f_{\mathbf{m}}(\mathbf{n})=\kappa\left[(\mathbf{m}-\mathbf{n})^2+1\right]^{-3/2}$ ($\kappa$ is a constant).

\section{Results}

Using the CTQMC algorithm \cite{kornilovitch}, we compute the splitting between $s$ and $p$ bipolaron states with the estimator $E_{sp}=-\ln (\langle s\rangle)/\beta$. In the $p$-state, $s=-1$ for exchanged, 1 for non exchanged configurations, and $s=0$ for paths with cohabiting endpoints. Inverse effective mass is calculated as $m_0/m^{**}=\langle s(\Delta r)^{2}\rangle/(\beta\hbar^2\langle s\rangle)$, and $\Delta \mathbf{r}$ is the twist in boundary conditions on the imaginary time axis.  The bipolaron action \cite{condmat} extends the polaron action \cite{kornilovitch}. We present triplet splitting in the Coulomb-Fr\"ohlich model. Below a certain el-ph coupling, the singlet-triplet pairing is degenerate, as Coulomb repulsion stops pairing. The triplet bipolaron is always heavier than the singlet state within numerical accuracy.
%\begin{figure}
%\begin{center}
%\includegraphics[width=55mm]{CrabCuprate.eps}
%\end{center}
%\caption{Crab bipolaron moves in the 1st order of the n.n.n.hopping, $t'$.}
%\end{figure}

\begin{figure}
\includegraphics[height=75mm,angle=270]{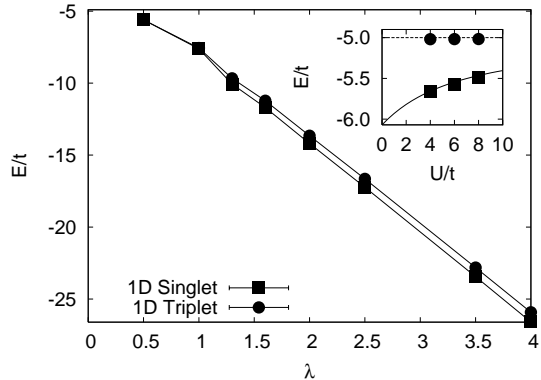}
\caption{Total energy of the CF model in 1D, $U/t=10$, $\omega/t=4$, $V=0$. (Inset: UV model in 1D with $V=4$. QMC results agree with analytic results \cite{kornilovitch2004a} as a test of the method).  El-ph coupling is defined as
$\lambda=\sum_{\mathbf{m}}f^{2}_{\mathbf{m}}(0)/2M\omega^2 zt$.}
\end{figure}

\begin{figure}
\includegraphics[height=75mm,angle=270]{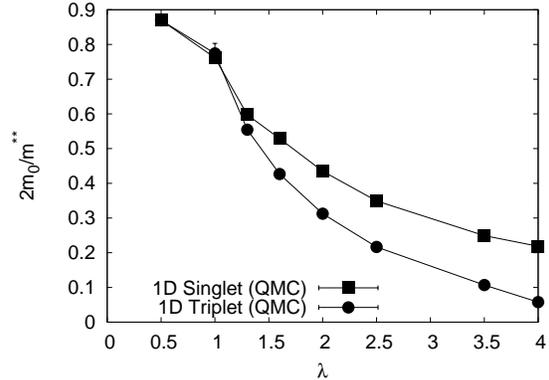}
\caption{Bipolaron inverse effective mass of the CF model in 1D. $U/t=10$, $V=0$ and $\omega/t=4$. N.B. Triplets are heavier than singlets, as in the solution of the UV model.}
\end{figure}

\section{Outlook for $d$-wave pairing}

We have used an advanced Quantum-Monte-Carlo algorithm to simulate bipolarons in the presence of very strong Coulomb repulsion with long-range el-ph interaction. We have used this algorithm to investigate the total energy of possible bipolaron pairing symmetries for the first time.

Given experimental evidence for a $d$-component in the cuprates, we aim to simulate the $d$-bipolaron as a prelude to multipolaron calculations.  Such as state has recently been seen in the two dimensional Holstein model as a consequence of large phonon frequency in an extended Eliashberg theory \cite{hague2006a}.
 For the $d$-state, the estimator is similar to the $p$-state, but now parallel end configurations contribute sign 1 and those rotated by $90^{o}$ sign -1 \cite{macridin}. New update rules are required in the d-state.  The bipolaron action for rotated boundary conditions in time is complicated, and will be discussed later.
While it is clear from the estimator for $E_{sd}$ that a simple bipolaron pair cannot have pure $d$-symmetry, when Coulomb repulsion is large, $d$ and $s$-wave paired states can be close in energy. Also, multipolaron effects may be capable of inducing a $d$-wave state \cite{alexandrov1998a,hague2006a}.

\section{Acknowledgements}

We acknowledge funding from EPSRC grant nos. EP/C518365/1 and no
EP/D07777X/1.

%\bibliographystyle{elsart-num}
%\bibliography{hague_proceedings}

% notes:

% \bibitem{label2}
% If there is a note, it should come last:
% \bibitem{label3} \note
% \end{subbibitems}

\end{document}